\documentclass[twocolumn]{aa}
\usepackage{graphicx}
\usepackage{txfonts}
\usepackage{graphicx}
\usepackage{natbib}
\usepackage{rotate}
\usepackage{epsf}
\begin{document}
%
\newcommand{\labeln}[1]{\label{#1}}
\newcommand{\Msolar}{M$_{\odot}$}
\newcommand{\Lsolar}{L$_{\odot}$}
\newcommand{\farcmin}{\hbox{$.\mkern-4mu^\prime$}}
\newcommand{\farcsec}{\hbox{$.\!\!^{\prime\prime}$}}
\newcommand{\kms}{\rm km\,s^{-1}}
\newcommand{\cc}{\rm cm^{-3}}
\newcommand{\Alfven}{$\rm Alfv\acute{e}n$}
\newcommand{\Vap}{V^\mathrm{P}_\mathrm{A}}
\newcommand{\Vat}{V^\mathrm{T}_\mathrm{A}}
\newcommand{\D}{\partial}
\newcommand{\DD}{\frac}
\newcommand{\TAW}{\tiny{\rm TAW}}
\newcommand{\mm }{\mathrm}
\newcommand{\Bp }{B_\mathrm{p}}
\newcommand{\Bpr }{B_\mathrm{r}}
\newcommand{\Bpz }{B_\mathrm{\theta}}
\newcommand{\Bt }{B_\mathrm{T}}
\newcommand{\Vp }{V_\mathrm{p}}
\newcommand{\Vpr }{V_\mathrm{r}}
\newcommand{\Vpz }{V_\mathrm{\theta}}
\newcommand{\Vt }{V_\mathrm{\varphi}}
\newcommand{\Ti }{T_\mathrm{i}}
\newcommand{\Te }{T_\mathrm{e}}
\newcommand{\rtr }{r_\mathrm{tr}}
\newcommand{\rbl }{r_\mathrm{BL}}
\newcommand{\rtrun }{r_\mathrm{trun}}
\newcommand{\thet }{\theta}
\newcommand{\thetd }{\theta_\mathrm{d}}
\newcommand{\thd }{\theta_d}
\newcommand{\thw }{\theta_W}
\newcommand{\beq}{\begin{equation}}
\newcommand{\eeq}{\end{equation}}
\newcommand{\ben}{\begin{enumerate}}
\newcommand{\een}{\end{enumerate}}
\newcommand{\bit}{\begin{itemize}}
\newcommand{\eit}{\end{itemize}}
\newcommand{\barr}{\begin{array}}
\newcommand{\earr}{\end{array}}
\newcommand{\eps}{\epsilon}
\newcommand{\veps}{\varepsilon}
\newcommand{\vepsdi}{{\cal E}^\mathrm{d}_\mathrm{i}}
\newcommand{\vepsde}{{\cal E}^\mathrm{d}_\mathrm{e}}
\newcommand{\ber}{\begin{array}}
\newcommand{\eer}{\end{array}}
\newcommand{\lraS}{\longmapsto}
\newcommand{\lra}{\longrightarrow}
\newcommand{\LRA}{\Longrightarrow}
\newcommand{\Equival}{\Longleftrightarrow}
\newcommand{\DRA}{\Downarrow}
\newcommand{\LLRA}{\Longleftrightarrow}
\newcommand{\diver}{\mbox{\,div}}
\newcommand{\grad}{\mbox{\,grad}}
\newcommand{\cd}{\!\cdot\!}
\newcommand{\Msun}{{\,{\cal M}_{\odot}}}
\newcommand{\Mstar}{{\,{\cal M}_{\star}}}
\newcommand{\Mdot}{{\,\dot{\cal M}}}
\title{ A method for relaxing the CFL-condition in time explicit schemes}
\bigskip\bigskip

\begin{raggedright}  
 
\author{ A. Hujeirat   }  
\institute{Max-Planck-Institut f\"ur Astronomie, 69117 Heidelberg, Germany}
\end{raggedright}
\offprints{A. Hujeirat, \email{hujeirat@mpia-hd.mpg.de}}

 \abstract{ A method for relaxing the CFL-condition, which limits the time step size
           in  explicit methods in computational fluid dynamics,  is presented.
           The method is based on re-formulating explicit methods in matrix form,
           and considering them as a special-Jacobi iteration scheme  
           that converge efficiently if the CFL- number is less than unity.
           By adopting this formulation, one can design various solution
           methods in arbitrary dimensions that range from  explicit to unconditionally stable implicit methods 
           in which  CFL-number could reach arbitrary large values. 
           In addition, we find that adopting a specially varying time stepping 
           scheme accelerates convergence toward steady state solutions and improves  
           the efficiently of the solution procedure.
     \keywords{Methods: numerical -- hydrodynamics -- MHD -- radiative transfer
               }
   }
\titlerunning{Relaxing the CFL-condition}
\maketitle 

\section{Introduction}
The majority of the numerical methods used in astrophysical fluid dynamics 
are based on time-explicit methods \cite[see][]{Stone, Ziegler, Koide}.\\
Advancing the solution in time in these methods is based on 
time-extrapolation procedures, that are found to be numerically stable if
the time-step size is shorter than a critical value, which is equivalent to the
requirement that the Courant-Friedrich-Levy (CFL) number must be smaller than unity. \\
This condition, however, limits the range of application and severely affects 
their robustness (see Fig. 1).
Using high performance  computers to perform a large number of explicit time steps
leads to accumulations of round-off errors that may easily distort the propagation of
information from the boundaries and cause divergence of the solution procedure,
 especially if Neumann type conditions are imposed at the boundaries. \\
In this paper we present for the first time a numerical strategy toward relaxing
the CFL-condition, and therefore enlarging the range of application of explicit methods.
\begin{figure*}[htb]
\begin{center}
{\hspace*{-0.5cm}
\includegraphics*[width=10.0cm,bb=0 0 505 52,clip]{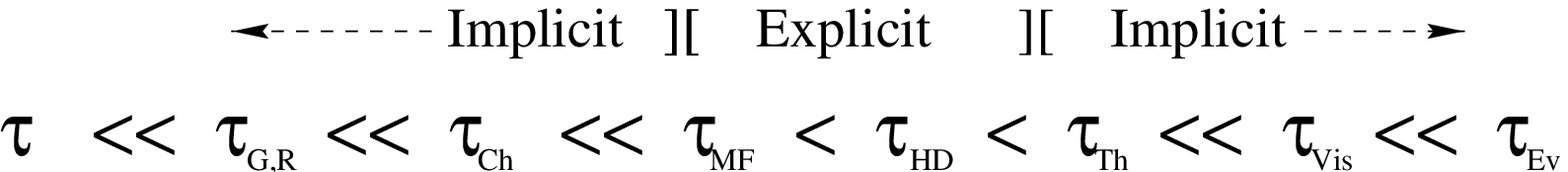}
}
\end{center}
{\vspace*{-0.4cm}}
\caption [ ] {      	The range of applicability of implicit and explicit methods.
	The equations describing physical and chemical processes
	whose characteristic time scales are much shorter than the
	hydrodynamic time scale ($\tau_{HD}$) should be solved implicitly.
	For example, the characteristic time corresponding to the
	propagation of gravitational waves ($\tau_G$), peaks in the radiative
	energy ($\tau_R$) and chemical processes ($\tau_{Ch}$)  occur on much
shorter time scales than 
	$\tau_{HD}.$ In most astrophysical problems {\Alfven} wave crossing time
	($\tau_{MHD}$) is shorter than $\tau_{HD}$.
	The time scale required for fluid flows (non-relativistic) to relax
	thermally ($\tau_{Th}$) is in general one order of magnitude
	larger than $\tau_{HD}$, depending on the efficiency of the cooling processes.
        However, the viscous time scale
($\tau_{Vis}$)
	can be significantly larger than $\tau_{HD}$ depending on how large the
	Reynolds number ($Re$) is. In most astrophysical flows, $Re$ is at
least $10^3$.
	If accretion is considered, the time scale for an envelope to evolve
	($\tau_{Ev}$) is approximately equal to the mass of the envelope
divided
	by the accretion rate. Thus, for an envelope of $\,{\cal M}_{En} =
10^{-5}
	{\cal M}_{\odot}\,$ to evolve from an accretion rate of
$\,\dot{\cal M}\sim
	10^{-10}{\cal M}_{\odot}/Y\,$, an accretion time of the order
$\,\tau_{Ev}
	\sim 10^5\,$ years is required.                 
  } 
\end{figure*}
\section{Mathematical formulation - scalar case}
In fluid flows, the equation of motion which describes the time-evolution of a quantity q in conservation form 
reads:
\beq
    \DD{\D q}{\D t} + L{q} \vec{V} = {f},
\eeq
where $\vec{V}$ and $f$ are the velocity field and external forces, respectively.
$L$ represents  a
first and/or second order spatial operator that describe the advection and diffusion of $q$. 
 
In the finite space $\cal{H}$, we may replace the time derivative of $q$ by:
\beq
  \DD{\delta q}{\delta t} =  \DD{q^\mm{n+1} - q^\mm{n}}{\delta t},
\eeq
where $q^\mm{n}$ and $q^\mm{n+1}$ correspond to the actual value of $q$ at the old and new time
levels, respectively. \\
An explicit formulation of Eq.1 reads: 
\beq
  \DD{\delta q}{\delta t} = [ -L{q} \vec{V} + {f}]^\mm{n},
\eeq
whereas the corresponding implicit form is:
\beq
  \DD{\delta q}{\delta t} = [ -L{q} \vec{V} + {f}]^\mm{n+1}.
\eeq
Combining these two approaches together, we obtain:
\beq
  \DD{\delta q}{\delta t} = \theta [ -L{q} \vec{V} + {f}]^\mm{n+1}
                          + (1-\theta) [ -L{q} \vec{V} + {f}]^\mm{n},
\eeq
where $\theta (0 \le \theta \le 1)$ is a switch on/off parameter.\\
To first order in $\delta t $, we may Taylor-expand $q^{n+1}$ and $f^{n+1}$ around $q^{n}$, i.e., 
$q^{n+1}=q^{n} + \delta q + \mathcal{O}(\delta^2)$, and obtain:
\beq
  \DD{\delta q}{\delta t}  + \theta (L \vec{V}  + {g})\delta q = RHS^\mm{n},
\eeq
where $g=\D f/\D q$ and  $RHS^\mm{n} =  [ {f}-L{q} \vec{V}]^\mm{n}$. 
In the following discussion, we term $RHS^\mm{n}$  as the time-independent
residual.\\
Applying Eq. 6 to the whole number of grid points, the following 
matrix-equation can be obtained:
\beq
  (\DD{I}{\delta t}  + \theta A)\delta q = RHS^\mm{n}.
\eeq
\begin{figure}[htb]
\begin{center}
{\hspace*{-0.5cm}
\includegraphics*[width=5.0cm,bb=0 0 310 345,clip]{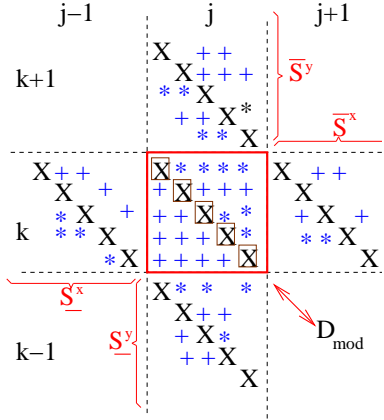}
}
\end{center}
{\vspace*{-0.4cm}}
\caption [ ] {The neighboring block matrices in the x and y-directions resulting
              from 5-star staggered grid  discretization. Entries marked with `X' denote
              the elements usually used in the implicit operator splitting approach 
              \cite{Hujeirat}, `*' and `+' are coefficients corresponding
              to the the source terms. The semi-explicit  method for a scalar equation
              relies on inverting the diagonal matrix whose entries are marked 
             with X surrounded by squares. The generalization of the
            semi-explicit  method to the multi-dimensional HD-equations requires inverting
            the  block diagonal matrix $D_\mm{mod}$.  
  } 
\end{figure}
The matrix $A$ contains coefficients such as $V/\Delta x$, $\eta/\Delta x^2$ and $g$ that
correspond to advection, diffusion and to the source terms, respectively. In terms of 
equation 7,  explicit methods are recovered by neglecting the matrix $\theta A$.
We note that 
since the switch parameter $\theta$ does not appear on the RHS of Eq. 7, explicit formulation does not directly
depend on $\theta$, but rather on the multiplication of $\theta A$. Thus,
for explicit method to converge, it is necessary that the matrix 
$(\DD{I}{\delta t}  + \theta A)$ is diagonally dominant, which implies that
$\theta A$ must be negligibly small.
In terms of matrix algebra, this means that the absolute value of the sum of elements in each raw
of $\theta A$ must be smaller than the corresponding diagonal element $1/{\delta t}$
\cite{Hackbusch}.
In the absence of diffusion and external forces (i.e., $\eta = f=0$),  this   
is equivalent to the requirement that at each grid point: $1/{\delta t} > |V|/\Delta x$, i.e., 
$CFL \equiv {\delta t}|V|/\Delta x < 1$.\\
On the other hand, the matrix $A$ can be decomposed as follows:
$ A = D + L + U$, where D is a matrix that consists of the diagonal elements
of $A$. L and U contain respectively the sub- and super-diagonal entries of $A$.
Noting that a conservative discretization of the 
advection-diffusion hydrodynamical equations (Navier-Stokes equations)
 gives rise to a $D$ that contains positive values, we may
reconstruct a modified diagonal matrix $ D_\mm{mod} = {I}/{\delta t} + \theta D$.
In terms of Eq. 7 we obtain:
\beq
   [D_\mm{mod}   + \theta(L + U)]\delta q = RHS^\mm{n}.
\eeq
A slightly  modified semi-explicit form can be obtained by neglecting the
entries of the matrix $\theta(L + U)$. In this case,  a necessary condition for the iteration
procedure to converge  is that the absolute value of the sum of elements in each raw
of $\theta(L + U) $ must be much smaller than the corresponding diagonal element
of $D_\mm{mod}$ in the same raw. In terms of Equation 8,   the method is said to converge
if the entries in each row of $D_\mm{mod}$ fulfill the following condition:
\beq
1/{\delta t} + \theta (|V|/\Delta x + \eta/\Delta x^2 + g ) > ||A-D||_\infty,
\eeq
where $||A-D||_\infty$ denotes the $\infty-$norm of $A-D$. This 
 can be achieved, however, if the flow is smooth, viscous,  and if appropriate  boundary conditions
are imposed\footnote{Note that diffusion pronounces the inequality in Eq. 9, which gives rise
to larger CFL-numbers.}. Consequently, the inversion process of $D_\mm{mod}\delta q = RHS^\mm{n}$ proceeds
stably even for large CFL-numbers (see Fig. 5).
\begin{figure}[htb]
\begin{center}
{\hspace*{-0.5cm}
\includegraphics*[width=6.75cm,bb=35 130 305 740 ,clip]{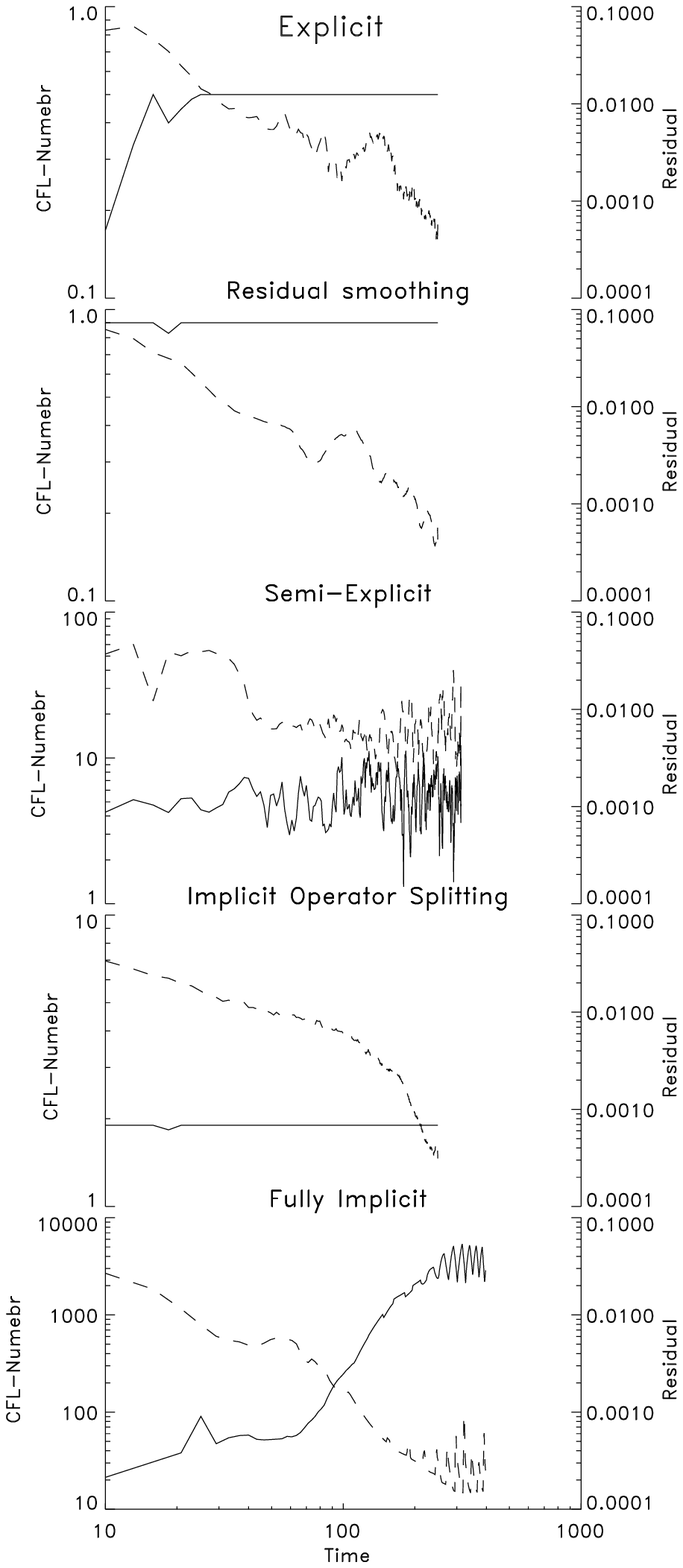}
}
\end{center}
{\vspace*{-0.4cm}}
\caption [ ] {      The development of the CFL-number (left axis) and the
           total residual (right axis) versus covered-time in normalized units of 
          five different numerical methods  (from to top to bottom: normal explicit,
            residual smoothing,
           semi-explicit, implicit operator splitting \cite{Hujeirat}
           and the fully implicit method). 
           While the effective time covered in each run of these different methods
           is similar, the actual number of iteration is substantially different.
           The numerical problem here is to search stationary solutions for
           Taylor flow between two concentric spheres.	
          The inner sphere has a radius $\,r_{in}\!=\!1\,$ and rotates with
  	angular velocity $\,\Omega_{in}\!=\!5\,$, whereas the outer sphere is 
         non-rotating and its radius is taken to be  $\,r_{out}\!=\!1.3\,$.
	We use the viscosity coefficient $\,\nu\!=\!10^{-2}$.
        The initial density and temperature are taken to be
       $\,\rho(r,\theta,t\!=\!0)\!=\!1\,$,
	and $\,T(r,\theta,t\!=\!0)\!=\!10^1\,$, respectively.
        The computational domain is 
	$\,[1,1.3]\times[0,\pi/2]\,$ and consists of   $\,30\!\times\!50\,$  non-uniformly distributed
	tensor-product mesh.\\                    
  } 
\end{figure}
\section{Generalization}
The set of 2D-hydrodynamical equations in conservative form and  in Cartesian coordinates 
 may be written in the following vector form:
\beq
 \DD{\D \vec{q}}{\D t} + L_\mm{x,xx} \vec{F} +   L_\mm{y,yy} \vec{G}  = \vec{f},
\eeq
where $F$ and $G$  are fluxes of $q$, and  
$ L_\mm{x,xx},\, L_\mm{y,yy}$ are first and second order transport operators
 that describe advection-diffusion
of  the vector variables $\vec{q}$ in x and y directions. $\vec{f}$ corresponds
to the vector of source functions.\\ 
By analogy with Eq.7, the linearization procedure applied to Eq. 10 
yields the following matrix form:
\beq
[\DD{I}{\delta t} + \theta(AL_\mm{x,xx} + B L_\mm{y,yy}-H)]\delta q = RHS^\mm{n},
\eeq
where $ A=\D F/\D q,\,  B=\D G/\D q$ and $ H= \D \vec{f}/\D q$, and which are evaluated
on the former time level. 
$RHS^\mm{n}=[\vec{f}- L_\mm{x,xx} \vec{F} -   L_\mm{y,yy} \vec{G}]^\mm{n}$.\\
Adopting a five star staggered grid discretization, it is easy to verify that
at each grid point Eq. 11 acquires the following block matrix equation:
\[
\DD{{\delta q}_\mm{j,k}}{\delta t} 
+ \underline{S}^\mm{x}{\delta q}_\mm{j-1,k} + {D}^\mm{x}{\delta q}_\mm{j,k}
                + \overline{S}^\mm{x}{\delta q}_\mm{j+1,k} \]
\beq
 + \underline{S}^\mm{y}{\delta q}_\mm{j,k-1} + {D}^\mm{y}{\delta q}_\mm{j,k}
                + \overline{S}^\mm{y}{\delta q}_\mm{j,k+1} 
= RHS^\mm{n}_\mm{j,k},
\eeq
where the underlines (overlines) mark the sub-diagonal (super-diagonal)
block matrices  in the corresponding directions (see Fig. 2).
 ${D}^\mm{x,y}$ are the
diagonal block matrices resulting from the discretization of the operators
$L_\mm{x,xx}\vec{F}$,  $L_\mm{y,yy} \vec{G}$ and $\vec{f}$.\\
To outline the directional dependence of the block matrices, we re-write Eq. 12 in a more compact form:
\beq
 \begin{array}{lll}
 & {\hspace*{0.3cm}}\overline{S}^\mm{y}{\delta q}_\mm{j,k+1} & \\
+ \underline{S}^\mm{x}{\delta q}_\mm{j-1,k} & + {D}_\mm{mod}{\delta q}_\mm{j,k} &
  + \overline{S}^\mm{x}{\delta q}_\mm{j+1,k}   = RHS^\mm{n}_\mm{j,k} \\
& + \underline{S}^\mm{y}{\delta q}_\mm{j,k-1}. & 
\end{array}
\eeq
where
\({D}_\mm{mod} = {{\delta q}_\mm{j,k}}/{\delta t} + {D}^\mm{x}+ {D}^\mm{y}.\) The
subscripts ``j'' and ``k'' denote the grid-numbering in the x and y directions, respectively
 (see Fig. 2). \\
\begin{figure}[htb]
\begin{center}
{\hspace*{-0.5cm}
\includegraphics*[width=6.5cm,bb=10 212 235 482,clip]{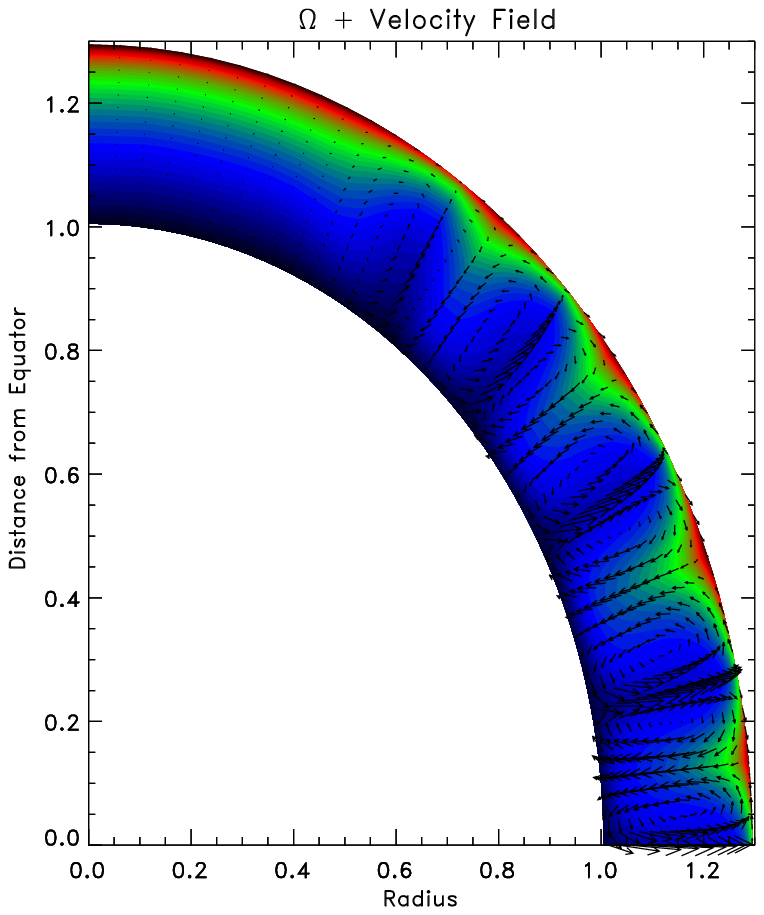}
}
\end{center}
{\vspace*{-0.4cm}}
\caption [ ] {Steady state solutions of the Taylor flow between two concentric spheres.
           Here the velocity field and the angular velocity 
          (large-to-low values correspond to blue-to-red colors) are shown.
	   The capability  of the methods to capture the formation of rotationally-driven 
           multiple vortices near-equatorial
           region is obvious.            
  } 
\end{figure}
Eq. 13 gives rise to three different types of solution procedures:
\ben
\item Classical explicit methods are obviously very special cases that are recovered
      if all the sub- and super-diagonal block matrices are neglected, as well as    
         ${D}^\mm{x}$ and ${D}^\mm{y}$. The only  matrix to be retained here
        is $ (1/{\delta t})\,\times\,$(the identity matrix), i.e.,
        the first term on the LHS of Eq. 12. This yields the vector equation:   
    \beq
        \DD{{\delta q}_\mm{j,k}}{\delta t} = RHS^\mm{n}_\mm{j,k}.
    \eeq
\item Semi-explicit methods are recovered when neglecting the sub- and super-diagonal
  block matrices only, but retaining the block diagonal matrices. In this case
      the matrix equation reads:
    \beq
       {D}_\mm{mod}{\delta q}_\mm{j,k} = RHS^\mm{n}_\mm{j,k}.
    \eeq
       We note that inverting ${D}_\mm{mod}$ is a straightforward procedure, either
       analytically or numerically. 
\item A fully implicit solution procedure requires retaining  
       all the  block matrices on the LHS of Eq. 13. This yields a global matrix that
       is highly sparse.  In this case, the 
      ``Approximate Factorization Method'' \cite[-AFM:][]{Beam} and the
     ``Line Gauss-Seidel Relaxation Method''
       \cite[-LGS:][]{MacCormack} are considered to be efficient solvers for the set of equations
        in multi-dimensions. 
\een
Fig. 3 shows the time-development of the CFL-number and the total residual 
for 5-different solution procedures for searching steady state solutions
for Taylor flows  between two concentric spheres. Using spherical geometry,
the set of the 2D axi-symmetric Navier-Stokes equations  are solved. The set consists of
 the three momentum equations, the continuity and the internal energy equations. The flow is assumed to
be adiabatic.\\
 In the explicit case, the equations are solved according to Eq. 14. 
For the semi-explicit procedure, we solve the HD-equations using the block matrix
formulation as described in Equation 15. The implicit operator splitting 
approach is based in solving each of the HD-equations implicitly. Here the
LGS-method is used in the inversion procedure of each equation \cite{Hujeirat}.
Unfortunately, while this method has been proven to be robust for modeling
compressible flows with open boundaries, it fails to achieve large CFL-numbers in
weakly incompressible flows (Fig. 3). This indicates that pressure
gradients in weakly incompressible flows do not admit splitting, and therefore they should
be included in the solution procedure simultaneously on the new time level. \\ 
In the final case, the whole set of HD-equations taking into account all pressure terms is solved in a fully implicit 
manner (Fig. 3/bottom). Here we use the  AFM for solving the general matrix-equation which is locally described by Eq. 13. \\
For constructing the time step size in these calculations, we have adopted the following
description: 
$\delta t = \alpha_0\eps/\max(RHS_{j,k})$, where $\eps = {\min}(\Delta X_j, \Delta Y_k),$ 
and where $\alpha_0$ is a constant of order unity. The maximum and minimum functions here run over
the whole number of grid points. 
\begin{figure}[htb]
\begin{center}
{\hspace*{-0.5cm}
\includegraphics*[width=7.5cm,bb=45 182 300 318,clip]{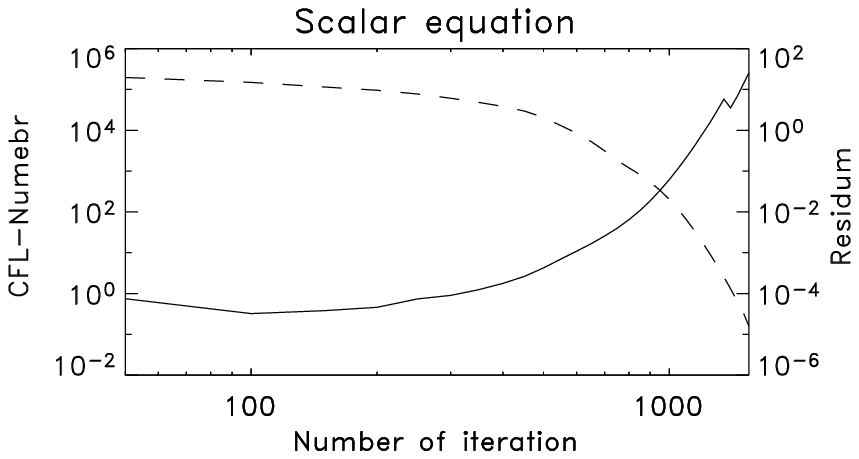}
}
\end{center}
{\vspace*{-0.4cm}}
\caption [ ] {The semi-explicit method applied to a scalar problem.
             The Figure shows the development of the CFL-number (left axis) and the 
           residual (right axis) of the angular momentum equation in two dimensions versus
            the iteration number. As initial conditions we use 
           the steady distributions of the physical variables that have been obtained
            from the simulations of the Taylor problem (see Fig. 4). This includes 
            the velocity field, density, temperature and $\eta$. For the 
            angular velocity we use $\Omega=0$ as initial condition.               
  } 
\end{figure}
\subsection{The specially varying time stepping scheme for accelerating convergence}
     Let [a,b] be the interval on which Eq. 1 is to be solved. We may divide [a,b] into
    N equally spaced finite volume cells: $\Delta x_\mm{i} = (b-a)/N$, $i=1,\, N$. To follow
    the time-evolution of q using a classical explicit method, the time step size
    must fulfill  the CFL-condition, which requires ${\delta t}$ to be smaller than the 
    critical value:
    $ {\delta t}^\mm{u}_\mm{c} = {\min}\{\Delta x_\mm{i}/(V  + V_\mm{S})_\mm{i}\}$.\\ 
    If [a,b] is divided into N highly stretched finite volume cells,
    for example $ \Delta x_\mm{1} < \Delta x_\mm{2} ... < \Delta x_\mm{N}$, then the CFL-condition
    restrict the time step size to be smaller than 
    $ {\delta t}^\mm{nu}_\mm{c} = {\min_i}\{(\Delta x_\mm{i}/(V + V_\mm{S})_\mm{i}\}$, which
    is much smaller than ${\delta t}^\mm{u}_\mm{c}$.  Thus, 
    applying a conditionally stable method to model flows using a highly non-uniform
    distributed mesh has the disadvantage that the time evolution of the variables in the whole domain 
    are artificially and severely affected by the flow behaviour on the finest cells.\\
    Moreover, advancing the variables in time may stagnate if the flow is strongly or nearly 
    incompressible.
    In this case, $V_\mm{S} >>  V,$ which implies that the time step size allowed by the CFL-condition
    approaches zero. \\
    However, we may  associate still  a time step size with each grid point, e.g., 
    $ {\delta t}^\mm{nu}_\mm{i} = \Delta x_\mm{i}/(V + V_\mm{S})$,  and follow the time
    evolution of each variable $q_\mm{i}$ independently. Interactions between variables
     enter the solution procedure through the evaluation of the 
     spatial operators on the former time level.
    This method, which is occasionally  called the ``Residual Smoothing Method'' proved to be efficient
    at providing quasi-stationary solutions within a reasonable number of iteration, when compared to
    normal explicit methods.
    \cite[Fig. 3, also see][]{Enander}. 
The main disadvantage of this method is its inability to provide physically meaningful
time scales for features that possess quasi-stationary behaviour.
Here we suggest to use the obtained quasi-stationary
solutions as initial configuration and re-start the calculations using
a uniform and physically well defined  time step size.
\section{Summary}
In this letter we have presented a strategy for relaxing the CFL-condition which
enlarges the range of application of explicit methods and improve their robustness.
The method is based on re-formulating explicit methods in matrix-form,
which can be then gradually modified up to a fully implicit scheme.
The matrix corresponding to the semi-explicit scheme presented here is a block diagonal
matrix that can be easily inverted, either analytically or numerically. Unlike normal
explicit methods, in which the inclusion of diffusion limits further the time step size,
in the semi-explicit formulation presented here diffusion
pronounces the diagonal dominance and enhances the stability of the inversion procedure, 
irrespective of the dimensionality of the problem.
We note that the  CFL-numbers achieved in the present modeling of Taylor flows  are, indeed, 
 larger than unity, but they are not impressively large  as we have  predicted theoretically. 
We may attribute this inconsistency to three different effects:
 1) The flow considered here is weakly incompressible. This means that the acoustic perturbations have the
largest propagation speeds, which requires that all pressure effects should be included in the
solution procedure simultaneously on the new time level.
  2) The conditions imposed on the
boundaries are non-absorbing, and do not permit advection of errors into regions exterior 
to the domain of calculations.
3) The method requires probably additional improvements in order to achieve large
CFL-numbers. This could be done, for example, within the context of the ``defect-correction' 
iteration procedure, in which the block diagonal matrix $D_\mm{mod}$  is employed as a pre-conditioner.
 
Finally, we have shown that the residual smoothing approach improves the 
convergence of explicit methods.  


\end{document}